\documentclass[10pt]{article}
\usepackage{graphicx}
\usepackage{amsmath}

\def\Title#1{\begin{center} {\Large #1 } \end{center}}
\def\Author#1{\begin{center}{ \sc #1} \end{center}}
\def\Address#1{\begin{center}{ \it #1} \end{center}}

\newcommand\pubblock{\rightline{\begin{tabular}{l} Proceedings of the Second Annual LHCP\\ \pubnumber\\
         \pubdate  \end{tabular}}}

\newenvironment{Abstract}{\begin{quotation} \begin{center} 
             \large ABSTRACT \end{center}\bigskip 
      \begin{center}\begin{large}}{\end{large}\end{center} \end{quotation}}

\newenvironment{Presented}{\begin{quotation} \begin{center} 
             PRESENTED AT\end{center}\bigskip 
      \begin{center}\begin{large}}{\end{large}\end{center} \end{quotation}}

\def\Acknowledgements{\bigskip  \bigskip \begin{center} \begin{large}
             \bf ACKNOWLEDGEMENTS \end{large}\end{center}}

\def\beq{\begin{equation}}
\def\eeq#1{\label{#1}\end{equation}}
\def\eeqn{\end{equation}}

\def\beqa{\begin{eqnarray}}
\def\eeqa#1{\label{#1}\end{eqnarray}}
\def\eeqan{\end{eqnarray}}

\let\bar=\overbar

\def\Dslash{\not{\hbox{\kern-4pt $D$}}}
\def\dslash{\not{\hbox{\kern-2pt $\del$}}}

\def\msb{{\bar{\ssstyle M \kern -1pt S}}}

\usepackage{color}

\newcommand\pubnumber{ }

\newcommand\pubdate{\today}

\def\affiliation{
On behalf of the LHCb collaboration, \\
Department of Physics \\
University of Warwick, Coventry, CV4 7AL, U.K. }

\begin{document}

\large
\begin{titlepage}
\pubblock

\vfill
\Title{ Searches for $\Lambda_{b}^{0}$ and $\Xi_{b}^{0}$ decays to $K_{\mathrm S}^{0}p\pi^{-}$ and 
$K_{\mathrm S}^{0}pK^{-}$ final states with the first observation of the $\Lambda_{b}^{0} \to K_{S}^{0}p\pi^{-}$ decay}
\vfill

\Author{ Mark Whitehead }
\Address{\affiliation}
\vfill
\begin{Abstract}

A search for previously unobserved decays of beauty baryons to the final states $K^0_{\mathrm{S}}p\pi^-$ 
and $K^0_{\mathrm{S}}pK^-$  is reported. The analysis is based on a data sample corresponding to an integrated 
luminosity of $1.0\,\mathrm{fb}^{-1}$ of $pp$ collisions. The $\Lambda^0_b \rightarrow K^0_{\mathrm{S}}p\pi^-$ decay 
is observed for the first time with a significance of $8.6\,\sigma$, and a measurement is made of the $CP$ asymmetry, 
which is consistent with zero. No significant signals are seen for  $\Lambda^0_b \rightarrow K^0_{\mathrm{S}}pK^-$ 
decays, $\Xi^0_b$ decays to both $K^0_{\mathrm{S}}p\pi^-$ and $K^0_{\mathrm{S}}pK^-$ final states, and the 
$\Lambda^0_b \rightarrow  D^-_s (K^0_{\mathrm{S}}K^-)p$ decay, and upper limits on their branching fractions are reported. 

\end{Abstract}
\vfill

\begin{Presented}
The Second Annual Conference\\
 on Large Hadron Collider Physics \\
Columbia University, New York, U.S.A \\ 
June 2-7, 2014
\end{Presented}
\vfill
\end{titlepage}
\def\thefootnote{\fnsymbol{footnote}}
\setcounter{footnote}{0}
%

\normalsize 


\section{Introduction}

Studies of $b$ baryon decays are at an early stage, with only a few known decay modes~\cite{Beringer:1900zz}. 
Most of these decays are of the 
$\Lambda^{0}_{b}$, the lightest ground state $b$ baryon. However, no three-body 
charmless hadronic final states had yet been observed. Conservation of baryon number means that searches for $CP$ 
violation do not require flavour tagging, which is challenging at hadron colliders. 
With large data samples in the future, Dalitz plot analyses of charmless three-body decays can provide 
further sensitivity to $CP$ violating observables.
The LHCb detector~\cite{Alves:2008zz} is a single arm spectrometer at the Large Hadron Collider. 
The data sample used corresponds to an integrated luminosity of $1.0~fb^{-1}$.
For more details about this analysis see Ref.~\cite{Aaij:2014lpa}.

\section{Selection}
The decay chain required is $\Lambda^0_b \rightarrow K^0_{\mathrm{S}}p h^-$ with 
$K^0_{\mathrm{S}} \to \pi^{+}\pi^{-}$, where $h$ is a kaon or pion.
Reconstruction of $K_{\mathrm S}^{0}$ candidates is split into two categories, 
{\it long} and {\it downstream} candidates. Long candidates are $K_{\mathrm S}^{0}$ 
candidates where the daughter tracks have hits in the 
vertex detector and tracking stations. Downstream candidates occur when the $K_{\mathrm S}^{0}$ 
daughters do not have hits in the vertex detector.
Boosted decision trees (BDTs) are used to separated signal decays from  
combinatorial background. The BDTs are trained separately for long and downstream 
candidates, using a simulated signal sample and high $B$ mass 
sideband data for the combinatorial background. 
Particle identification requirements are applied to discriminate between pions, kaons and protons. 
Charmless decays ($\Lambda_{b}^{0} (\Xi_{b}^{0}) \to K_{\mathrm S}^{0} ph^{-}$) 
are separated from their charmed counterparts 
($\Lambda_{b}^{0} \to \Lambda_{c}^{+} h^{-} (D_{s}^{-}p)$).
The decay $B^{0} \to K_{\mathrm S}^{0}\pi^{+}\pi^{-}$ is used as a normalisation channel.

\section{Fitting}

\begin{figure}[!htb]
 \centering
 \includegraphics[scale=0.28]{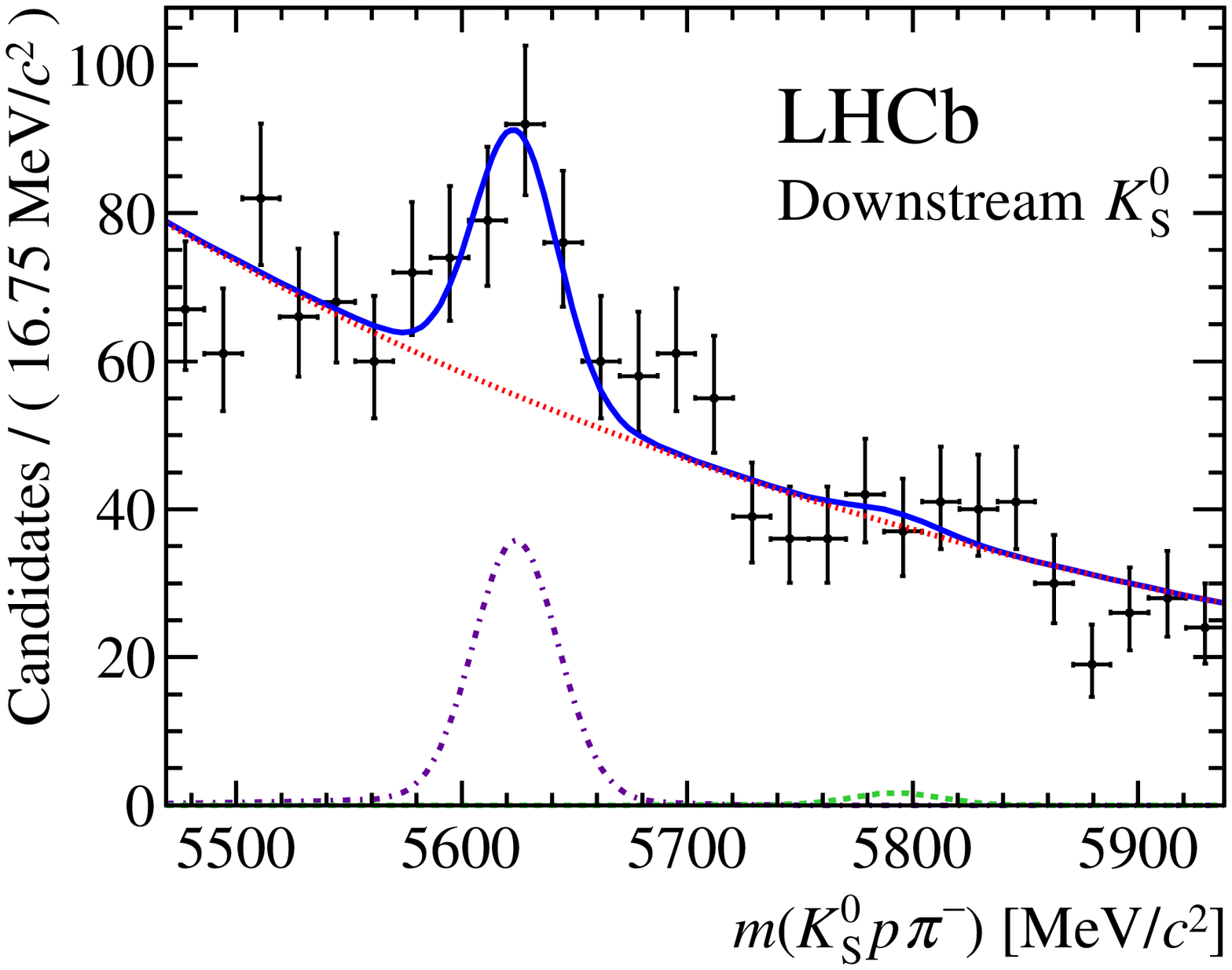}
 \includegraphics[scale=0.28]{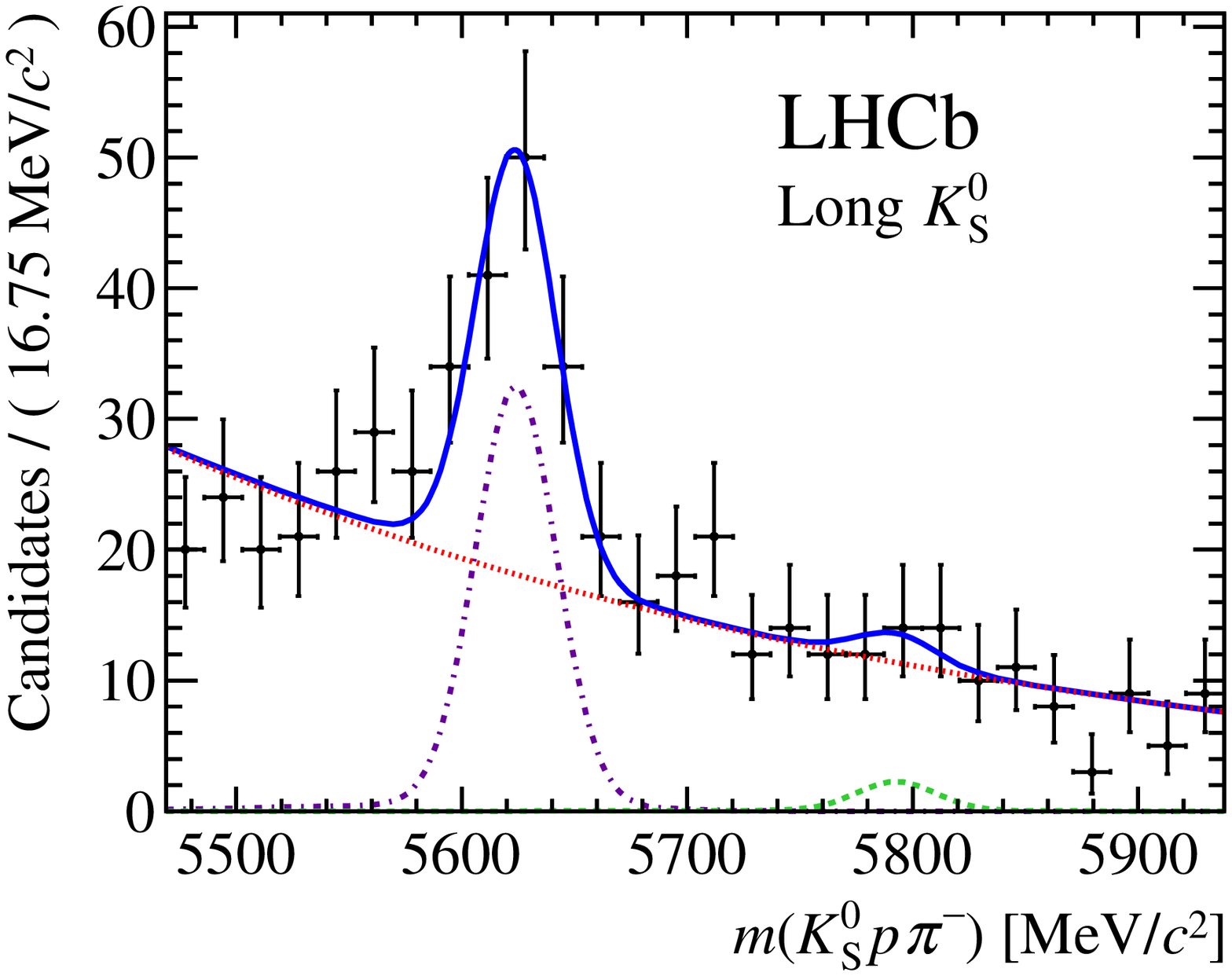}
 \includegraphics[scale=0.28]{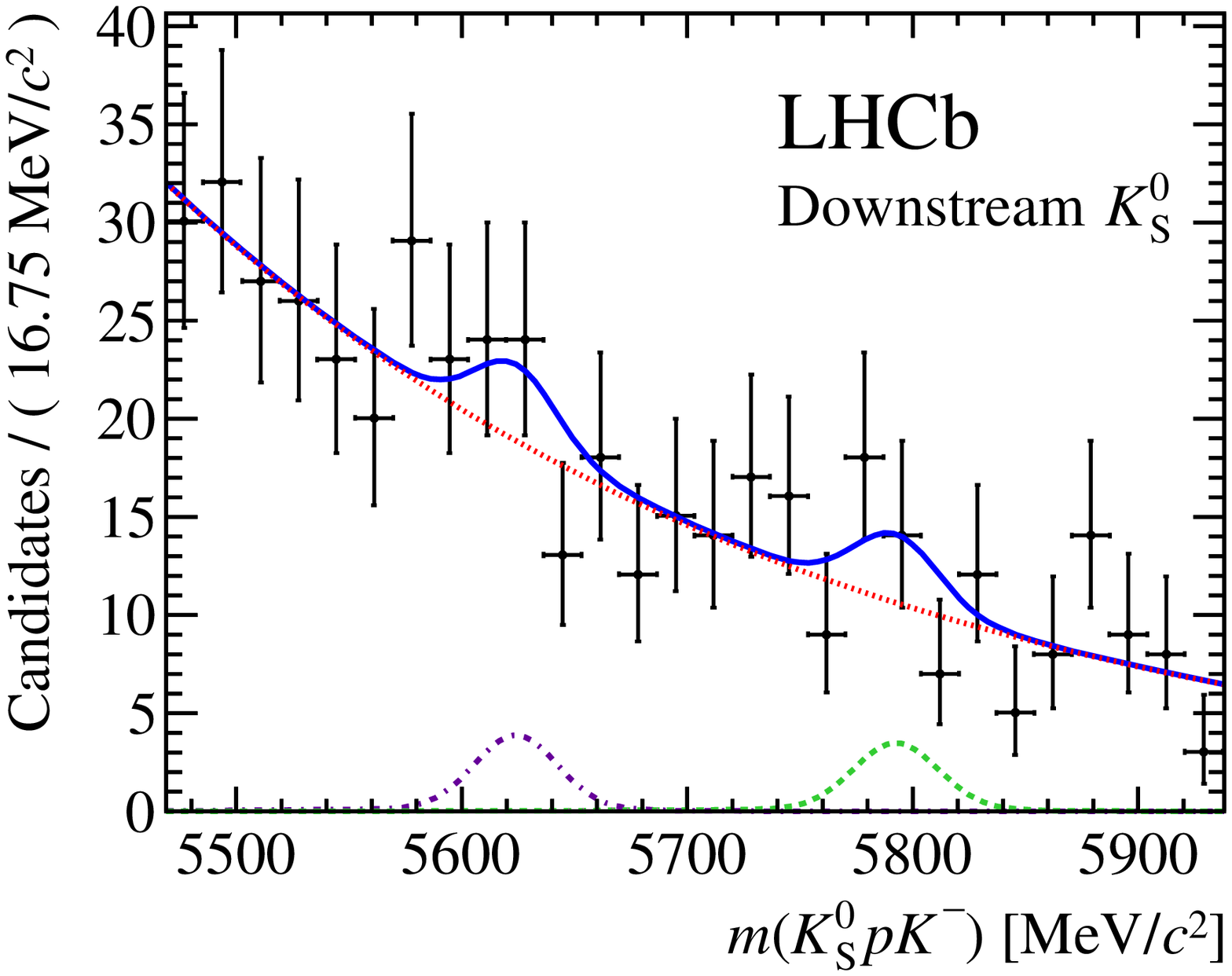}
 \includegraphics[scale=0.28]{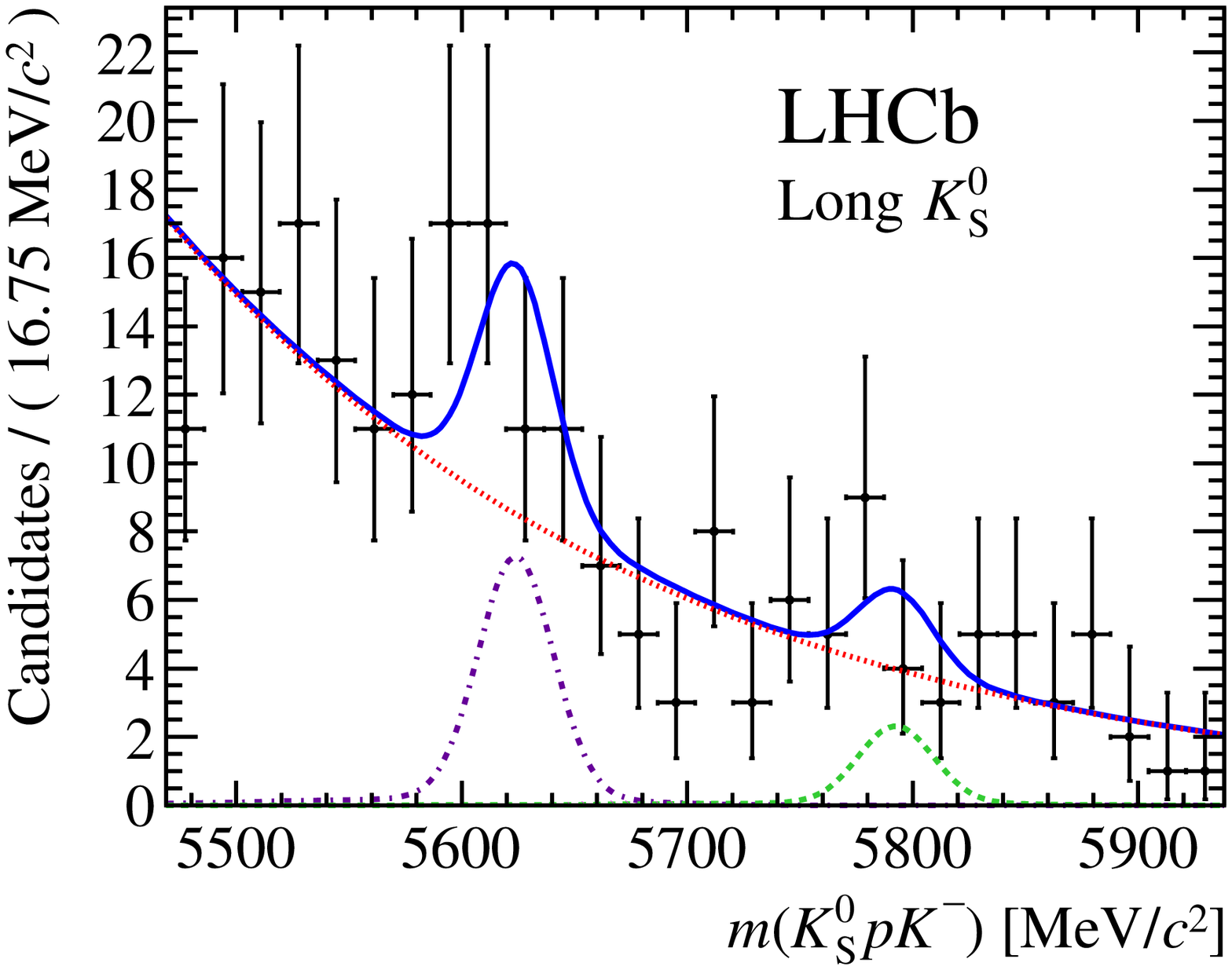}
 \caption{\small Fits to the $b$ baryon candidate invariant mass distribution for (top) $\Lambda_{b}^{0} \to K^{0}_{\mathrm S} p \pi^{-}$ 
and (bottom) $\Lambda_{b}^{0} \to K^{0}_{\mathrm S} p K^{-}$ decays. The left (right) column shows 
downstream (long) $K^{0}_{\mathrm S}$ candidates only. The visible shapes are the (blue) full fit, 
(violet) $\Lambda_{b}^{0}$ signal, (green) $\Xi^{0}_{b}$ signal and (red) combinatorial background.}
 \label{fig1}
\end{figure}

\begin{figure}[!htb]
 \centering
 \includegraphics[scale=0.25]{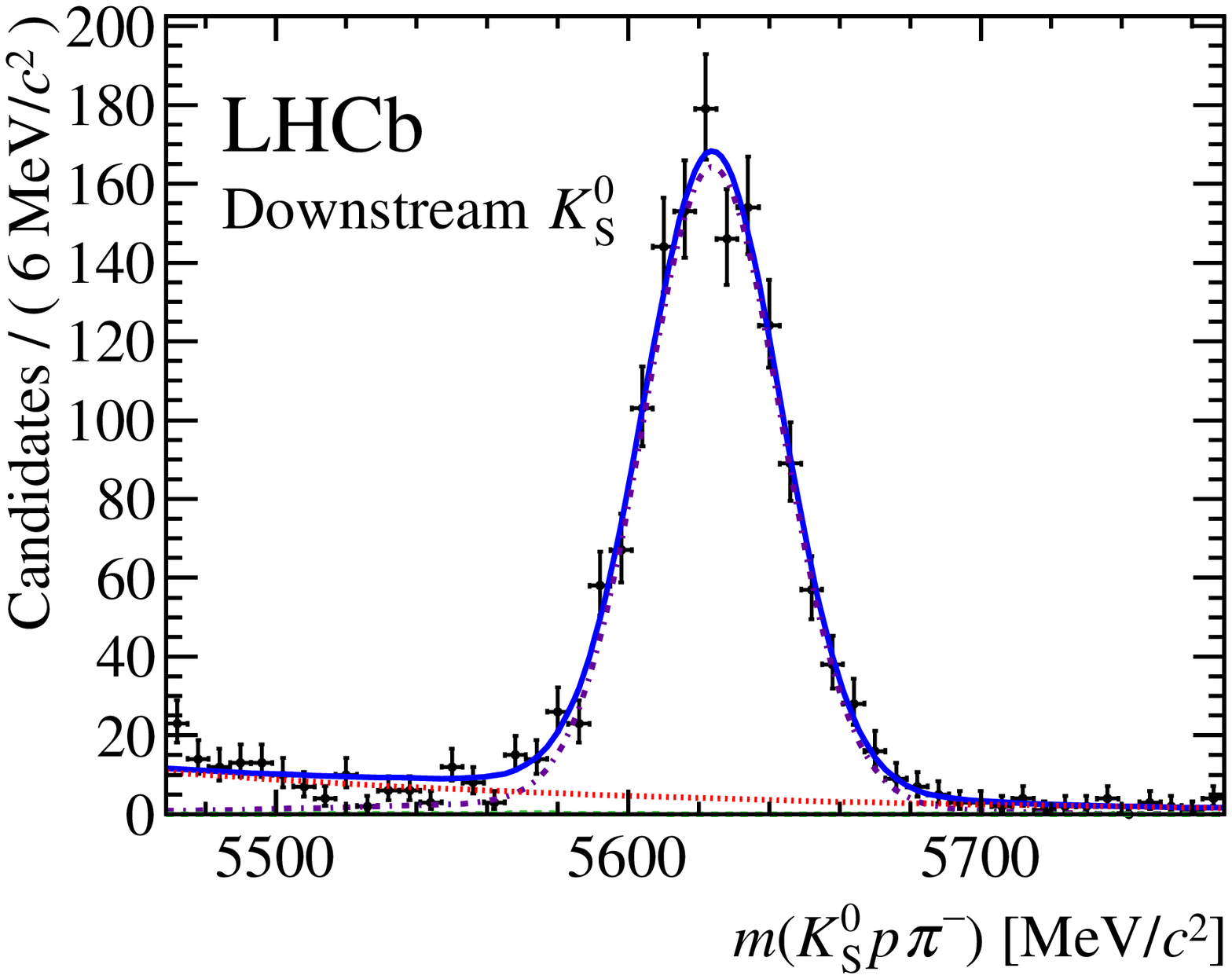}
 \includegraphics[scale=0.25]{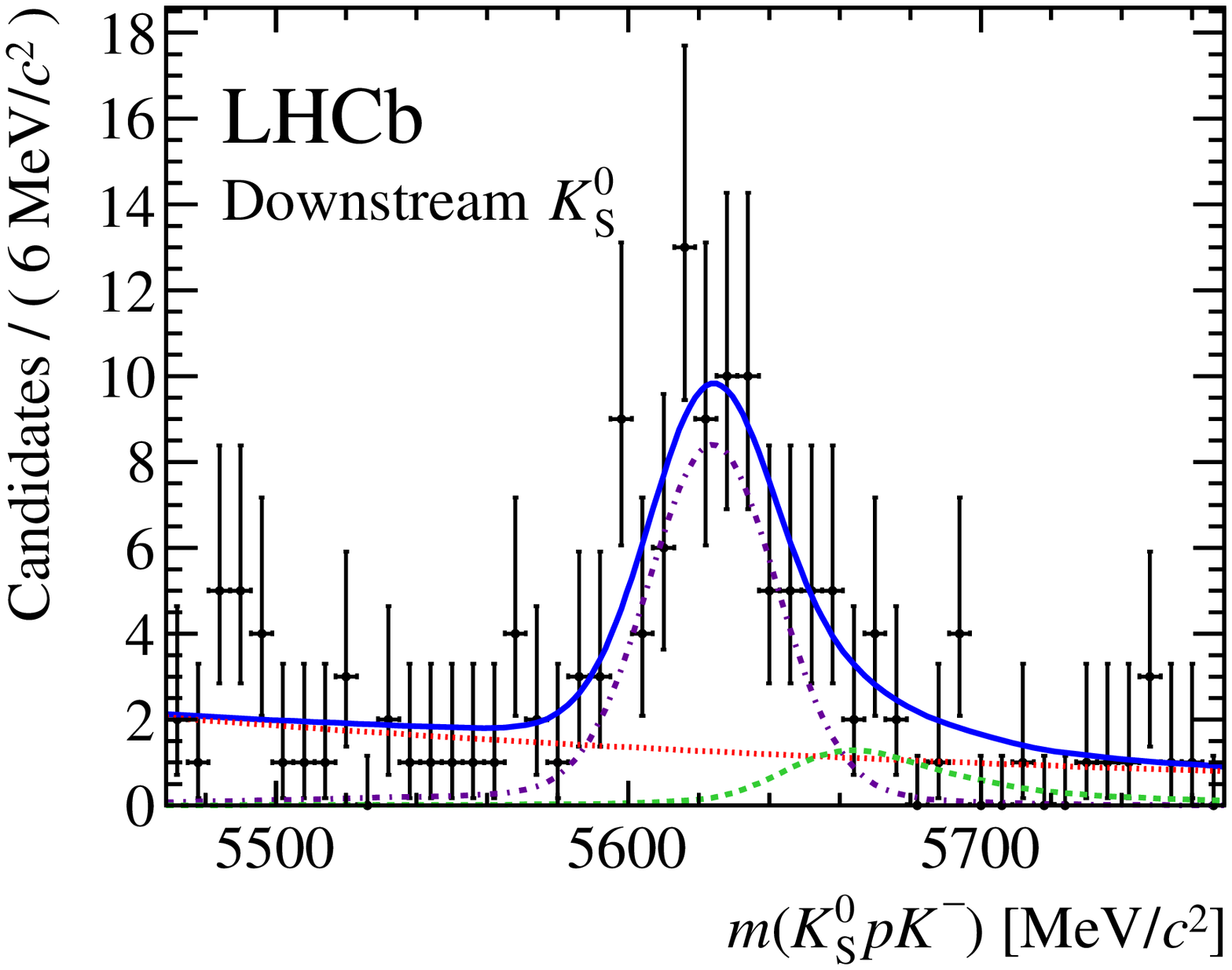}
 \includegraphics[scale=0.25]{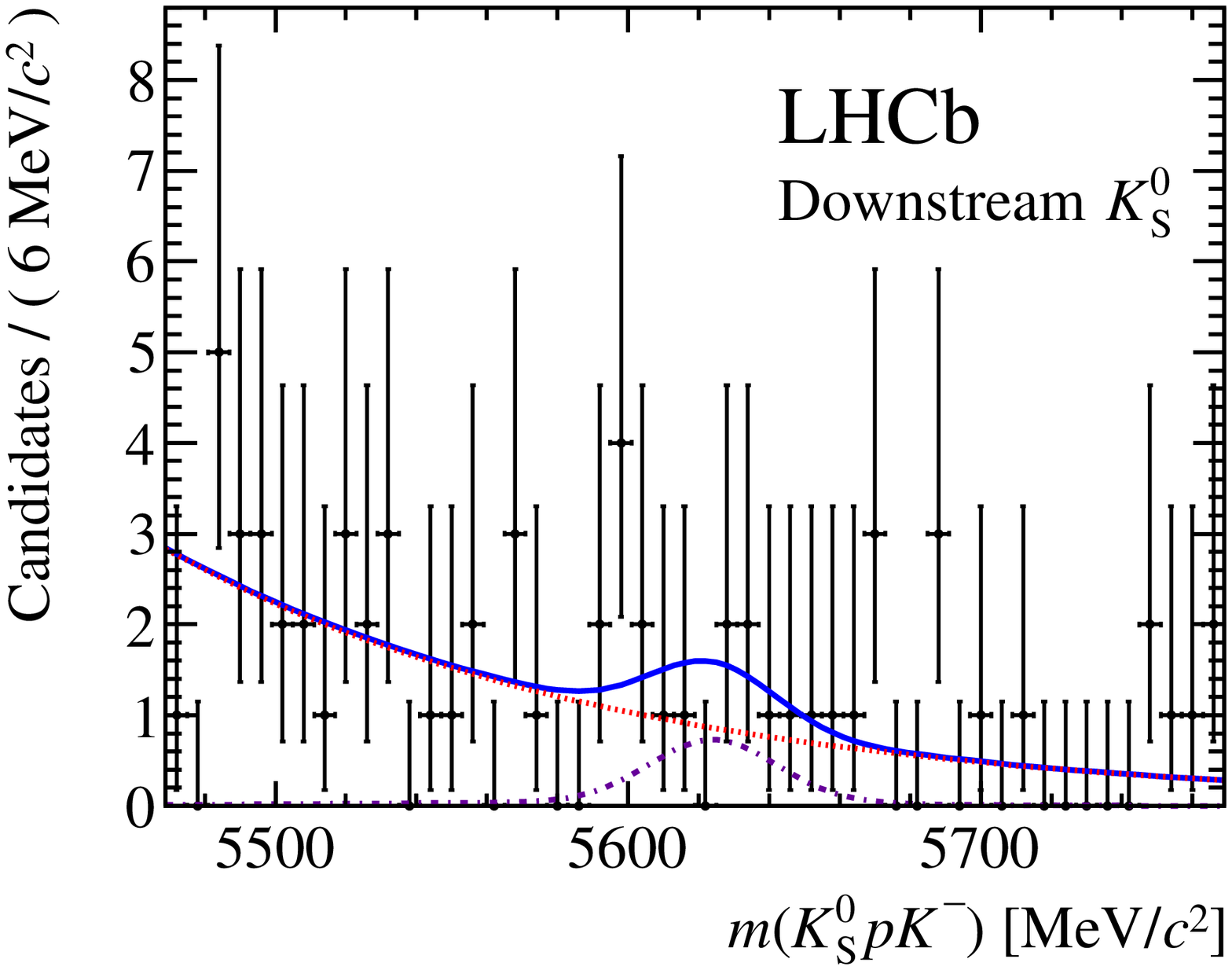}
 \includegraphics[scale=0.25]{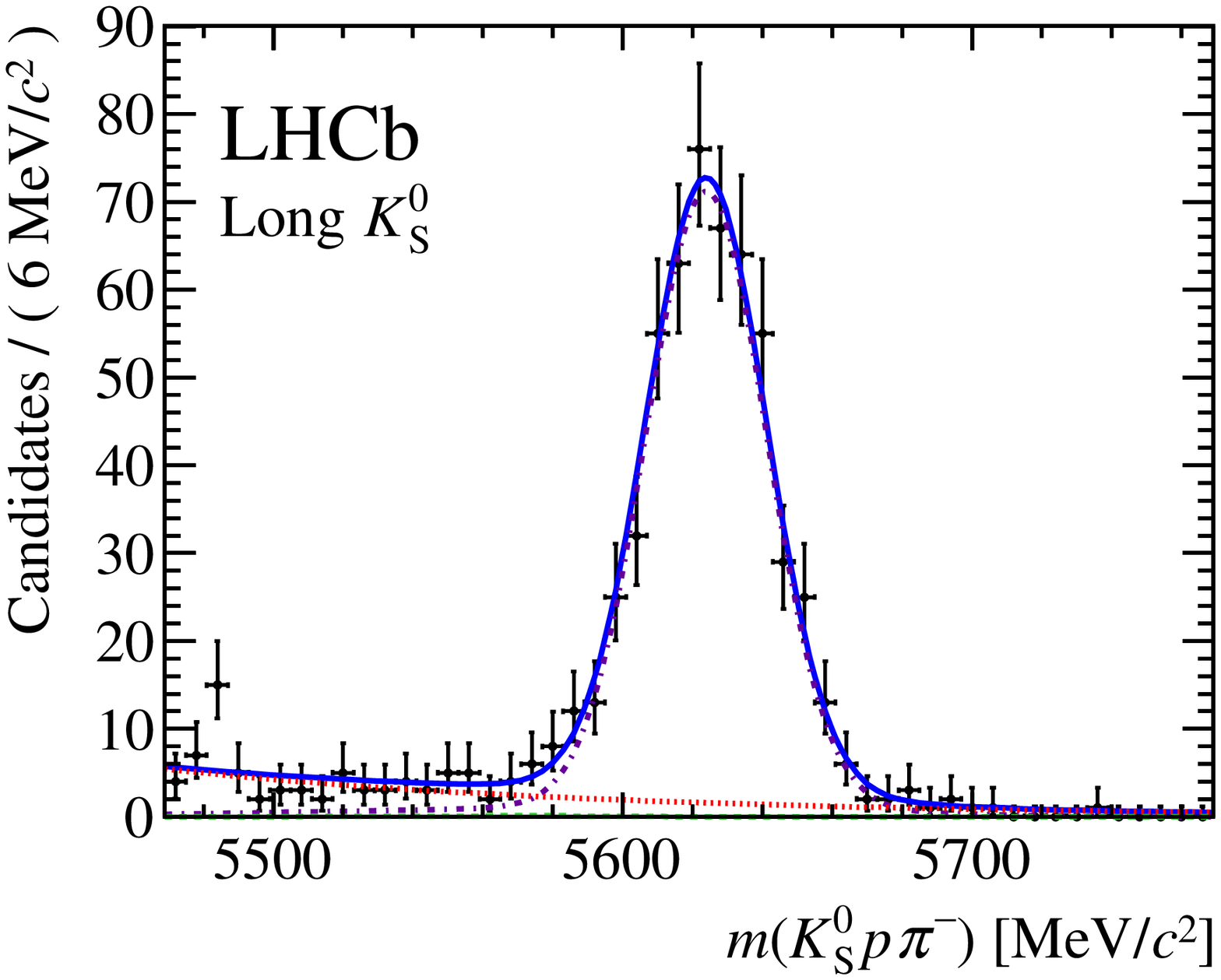}
 \includegraphics[scale=0.25]{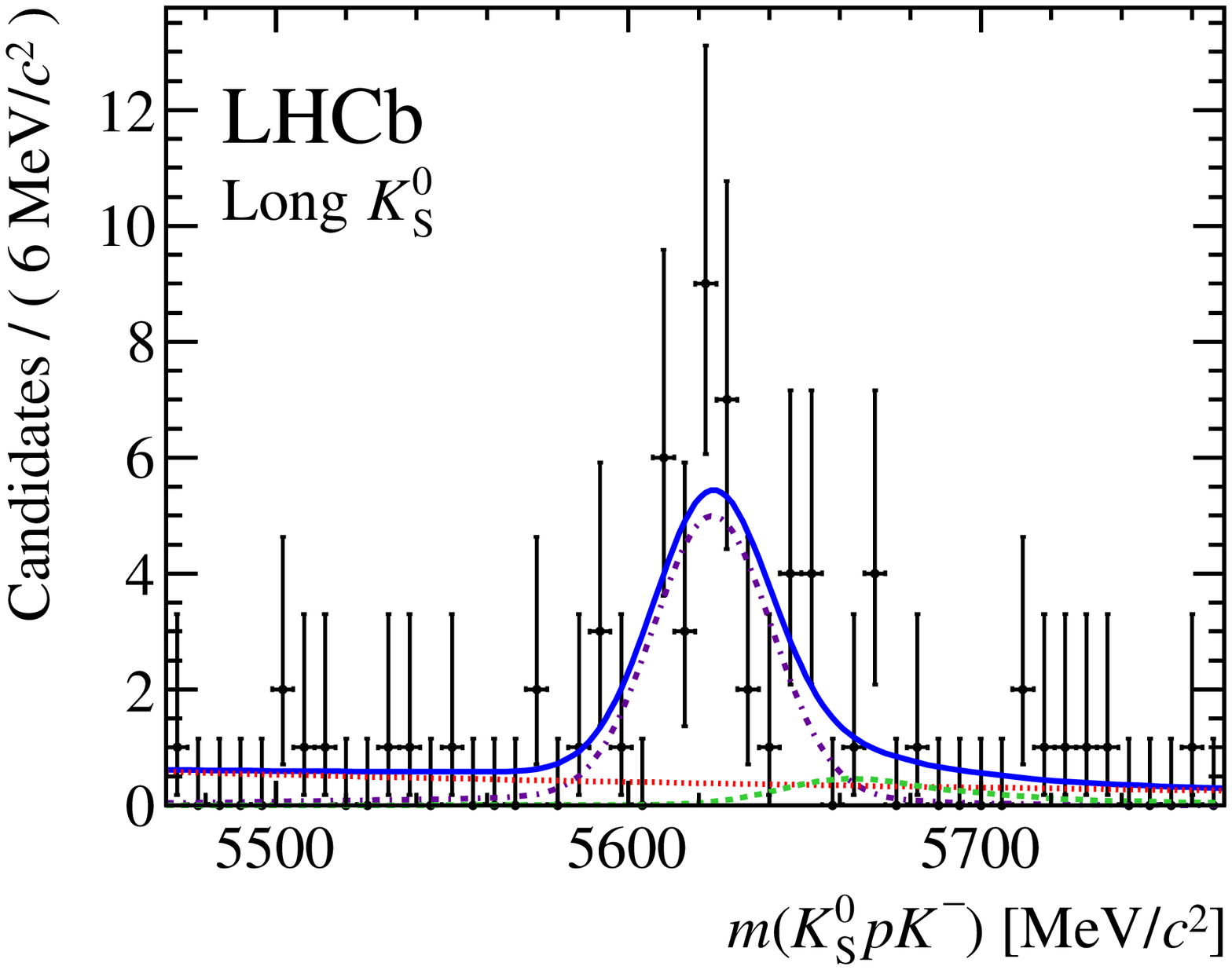}
 \includegraphics[scale=0.25]{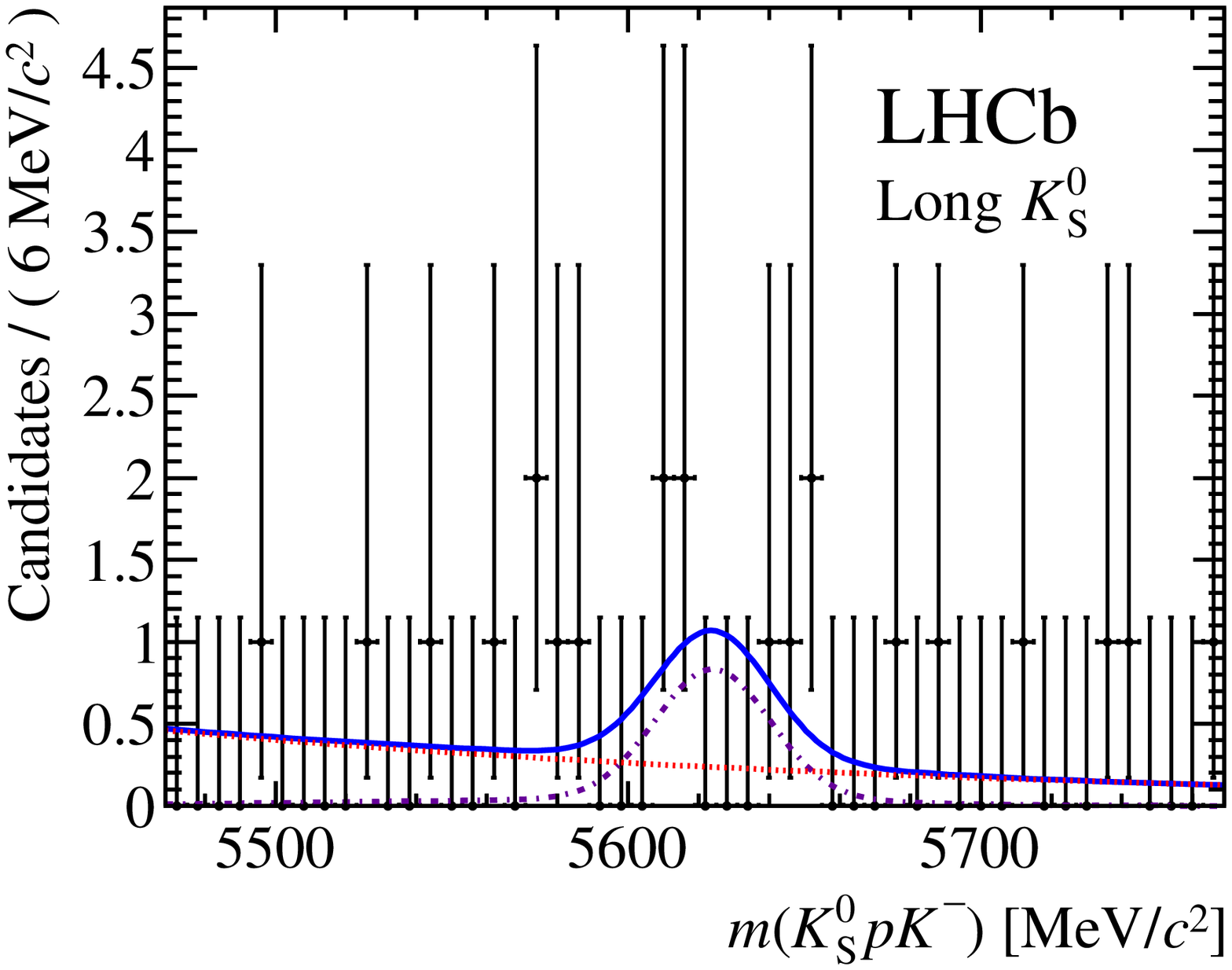}
 \caption{\small Fits to the $b$ baryon candidate invariant mass distribution for (left) 
$\Lambda_{b}^{0} \to \Lambda^{+}_{c} \pi^{-}$, (middle) $\Lambda_{b}^{0} \to \Lambda^{+}_{c} K^{-}$ and 
(right) $\Lambda_{b}^{0} \to D^{-}_{s} p$ decays. The top (bottom) column shows downstream (long) 
$K^{0}_{\mathrm S}$ candidates only. The visible shapes are the (blue) full fit, 
(violet) $\Lambda_{b}^{0}$ signal, (green) cross-feed contributions and (red) combinatorial background.}
 \label{fig2}
\end{figure}

The data is fitted using an extended unbinned maximum likelihood fit to the $b$ baryon candidate
invariant mass distribution, performed simultaneously to all decay modes. 
Signal peaks are modelled with 
the sum of a Gaussian and a bifurcated Gaussian, the combinatorial background is fitted 
by an exponential shape and the mis-identified backgrounds
with a double Crystal Ball~\cite{Skwarnicki:1986xj} function determined from simulation.
The mass difference between the $\Lambda^{0}_{b}$ and $\Xi^{0}_{b}$ baryons is fixed 
to $168.6\pm5.0\, {\rm MeV}/c^{2}$~\cite{Beringer:1900zz}.
The fit is shown in Figs.~\ref{fig1} and~\ref{fig2} for the charmless and 
charm modes, respectively. The signal yields are summarised in Table~\ref{tab1}.

\begin{table}[!htb]
 \centering
 \caption{\small Yields of the various decay modes from the simultaneous fit with statistical uncertainties.}
 \label{tab1}
 \begin{tabular}{ccc}
 \hline
 Decay mode & Downstream yield & Long yield \\
 \hline
 $\Lambda_{b}^{0} \to K^{0}_{\mathrm S} p \pi^{-}$ & $106.1 \pm 21.5$ & $90.0 \pm 14.6$  \\
 $\Lambda_{b}^{0} \to K^{0}_{\mathrm S} p K^{-}$ & $11.5 \pm 10.7$ & $19.6 \pm 8.5$  \\  
 $\Xi_{b}^{0} \to K^{0}_{\mathrm S} p \pi^{-}$ & $5.3 \pm 15.7$ & $6.4 \pm 8.5$  \\
 $\Xi_{b}^{0} \to K^{0}_{\mathrm S} p K^{-}$ & $10.5 \pm 8.8$ & $6.3 \pm 5.6$  \\
 \hline
 $\Lambda_{b}^{0} \to \Lambda^{+}_{c}(\to p\bar{K}^{0}) \pi^{-}$ & $1391.6 \pm 39.6$ & $536.8 \pm 24.6$  \\
 $\Lambda_{b}^{0} \to \Lambda^{+}_{c}(\to p\bar{K}^{0}) K^{-}$ & $70.0 \pm 10.3$ & $37.4 \pm 7.1$  \\
 $\Lambda_{b}^{0} \to D^{-}_{s}(\to {K}^{0} K^{-}) p$ & $70.0 \pm 10.3$ & $37.4 \pm 7.1$  \\
 \hline
 \end{tabular}
\end{table}

\section{Results}

Branching fractions are determined relative to $\mathcal{B}(B^{0} \to K^{0}\pi^{+}\pi^{-})$ 
and the known value~\cite{Beringer:1900zz} is used to determine the absolute branching fractions. 
Efficiency corrections and corrections due to the fragmentation fraction ($f_{\Lambda_{b}^{0}}/f_{d}$) are applied.
The results are quoted for $K^{0}$ or $\bar{K}^{0}$, according to 
the expectation for each decay
\begin{align}
\centering 
\nonumber \mathcal{B}(\Lambda_{b}^{0} \to \bar{K}^{0} p \pi^{-}) &= (1.26 \pm 0.19 \pm 0.09 \pm 0.34 \pm 0.05) \times 10^{-5},\\ 
\nonumber \mathcal{B}(\Lambda_{b}^{0} \to {K}^{0} p K^{-}) &= (1.8 \pm 1.2 \pm 0.8 \pm 0.5 \pm 0.1) \times 10^{-6}, < 3.5\,(4.0) \times 10^{-6} \,{\rm at}\, 90\,\% \, (95\,\%) \,{\rm CL},\\
\nonumber f_{\Xi_{b}^{0}}/f_{d} \,\times\, \mathcal{B}(\Xi_{b}^{0} \to \bar{K}^{0} p \pi^{-}) &= (0.6 \pm 0.7 \pm 0.2) \times 10^{-6}, < 1.6\,(1.8) \times 10^{-6} \,{\rm at}\, 90\,\% \, (95\,\%) \,{\rm CL},\\
\nonumber f_{\Xi_{b}^{0}}/f_{d} \,\times\, \mathcal{B}(\Xi_{b}^{0} \to \bar{K}^{0} p K^{-}) &= (0.6 \pm 0.4 \pm 0.2) \times 10^{-6}, < 1.1\,(1.2) \times 10^{-6} \,{\rm at}\, 90\,\% \, (95\,\%) \,{\rm CL},\\
\nonumber \mathcal{B}(\Lambda_{b}^{0} \to \Lambda^{+}_{c}(\to p\bar{K}^{0}) \pi^{-}) &= (1.40 \pm 0.07 \pm 0.08 \pm 0.38 \pm 0.06) \times 10^{-4},\\
\nonumber \mathcal{B}(\Lambda_{b}^{0} \to \Lambda^{+}_{c}(\to p\bar{K}^{0}) K^{-}) &= (0.83 \pm 0.10 \pm 0.06 \pm 0.23 \pm 0.03) \times 10^{-5},\\
\nonumber \mathcal{B}(\Lambda_{b}^{0} \to D^{-}_{s}(\to {K}^{0} K^{-}) p) &= (2.0 \pm 1.1 \pm 0.2 \pm 0.5 \pm 0.1) \times 10^{-6}, < 3.5\,(3.9) \times 10^{-6} \,{\rm at}\, 90\,\% \, (95\,\%) \,{\rm CL}.
\end{align}
For $\Lambda^{0}_{b}$ decays the uncertainties are statistical, systematic, 
plus those arising from the external inputs $f_{\Lambda_{b}^{0}}/f_{d}$ and $\mathcal{B}(B^{0} \to K^{0}\pi^{+}\pi^{-})$, respectively. 
For $\Xi_{b}^{0}$ decays the fragmentation fraction 
is unknown and the uncertainty from $\mathcal{B}(B^{0} \to K^{0}\pi^{+}\pi^{-})$ is negligible.
The $\Lambda_{b}^{0} \to \Lambda^{+}_{c} h^{-}$ branching fractions can be 
determined more precisely since 
$\mathcal{B}(\Lambda^{+}_{c}\to p\bar{K}^{0})$/$\mathcal{B}(\Lambda^{+}_{c}\to p K^{-} \pi^{+})$ 
is better known than $\mathcal{B}(\Lambda^{+}_{c}\to p K^{-} \pi^{+})$, which dominates 
the uncertainty on $f_{\Lambda_{b}^{0}}/f_{d}$. Dividing the branching fractions by 
$\mathcal{B}(\Lambda^{+}_{c}\to p K^{-} \pi^{+})$ and the ratio of $\Lambda_{c}^{+}$ 
branching fractions gives
\begin{align}
\centering
\nonumber \mathcal{B}(\Lambda_{b}^{0} \to \Lambda^{+}_{c} \pi^{-}) &= (5.97 \pm 0.28 \pm 0.34 \pm 0.70 \pm 0.24) \times 10^{-3},\\
\nonumber \mathcal{B}(\Lambda_{b}^{0} \to \Lambda^{+}_{c} K^{-}) &= (3.55 \pm 0.44 \pm 0.24 \pm 0.41 \pm 0.14) \times 10^{-4}.
\end{align}
The known value of $\mathcal{B}(D_{s}^{-}\to K^{0} K^{-})$ can be used to get
\begin{equation}
\centering
\nonumber
\mathcal{B}(\Lambda_{b}^{0} \to D^{-}_{s} p) = (2.7 \pm 1.4 \pm 0.7 \pm 0.1 \pm 0.1) \times 10^{-4}, < 4.8\,(5.3) \times 10^{-4} \,{\rm at}\, 90\,\% \, (95\,\%) \,{\rm CL},
\end{equation}
where the last uncertainty is from $\mathcal{B}(D_{s}^{-}\to K^{0} K^{-})$.

The large $\Lambda_{b}^{0} \to K^{0}_{\mathrm S} p \pi^{-}$ signal yield allows a 
measurement of the integrated $CP$ asymmetry, defined as 
\begin{equation}
\centering
\nonumber
\mathcal{A}_{CP}^{\rm RAW} = \frac{N_{\bar{f}} - N_{f}}{N_{\bar{f}} + N_{f}},
\end{equation}
where $N_{f(\bar{f})}$ is the signal yield for $\bar{\Lambda}_{b}^{0}$($\Lambda_{b}^{0}$) decays.
Corrections are applied for detection ($\mathcal{A_{\rm D}}$) and production 
($\mathcal{A_{\rm P}}$) asymmetries to give 
$\mathcal{A}_{CP} = \mathcal{A}_{CP}^{\rm RAW} - \mathcal{A_{\rm P}} - \mathcal{A_{\rm D}}$.
The corrections are taken from $\Lambda_{b}^{0} \to \Lambda^{+}_{c}(\to p\bar{K}^{0}) \pi^{-}$ 
decays, which are expected to have negligible $CP$ violation.
The corrected value is
\begin{equation}
\centering
\nonumber
\mathcal{A}_{CP}(\Lambda_{b}^{0} \to K^{0}_{\mathrm S} p \pi^{-}) = 0.22 \pm 0.13 (\rm stat) \pm 0.03 (\rm syst),
\end{equation}
which is consistent with zero.
The decay $\Lambda_{b}^{0} \to K^{0}_{\mathrm S} p \pi^{-}$ is observed for the first time with 
a significance of $8.6\,\sigma$, allowing a measurement of its integrated $CP$ asymmetry. 
Limits are set for the other decays modes where the signal yields 
are not significant. This work opens up exciting possibilities to study such decays with large 
data samples in the future.

\Acknowledgements
I thank the members of the LHCb collaboration for their help in preparing the poster and this 
document. Work supported by the European Research Council under FP7.

\end{document}